\documentclass[twocolumn,showpacs,superscriptaddress,amsmath,amssymb,aps,pra]{revtex4-1}
\usepackage{bm}
\usepackage{graphicx}
\usepackage{dcolumn}
\usepackage{amsmath,txfonts}
\usepackage{epstopdf}
\usepackage[mathlines]{lineno}
\usepackage{color}
\usepackage[colorlinks=true,linkcolor=blue,citecolor=blue]{hyperref}
\usepackage[normalem]{ulem}

\begin{document}

\title{Detection sensitivity enhancement of magnon Kerr nonlinearity in cavity magnonics induced by coherent perfect absorption}

\author{Guo-Qiang Zhang}
\email{zhangguoqiang@hznu.edu.cn}
\affiliation{School of Physics, Hangzhou Normal University, Hangzhou, Zhejiang 311121, China}

\author{Yimin Wang}
\email{vivhappyrom@163.com}
\affiliation{Communications Engineering College, Army Engineering University of PLA, Nanjing 210007, China}

\author{Wei Xiong}
\email{xiongweiphys@wzu.edu.cn}
\affiliation{Department of Physics, Wenzhou University, Zhejiang 325035, China}

\begin{abstract}
We show how to enhance the detection sensitivity of magnon Kerr nonlinearity (MKN) in cavity magnonics. The considered cavity-magnon system consists of a three-dimensional microwave cavity containing two yttrium iron garnet (YIG) spheres, where the two magnon modes (one has the MKN, while the other is linear) in YIG spheres are simultaneously coupled to microwave photons. To obtain the effective gain of the cavity mode, we feed two input fields into the cavity. By choosing appropriate parameters, the coherent perfect absorption of the two input fields occurs, and the cavity-magnon system can be described by an effective non-Hermitian Hamiltonian. Under the pseudo-Hermitian conditions, the effective Hamiltonian can host the third-order exceptional point (EP3), where the three eigenvalues of the Hamiltonian coalesce into one. When the magnon frequency shift $\Delta_K$ induced by the MKN is much smaller than the linewidths $\Gamma$ of the peaks in the transmission spectrum of the cavity (i.e., $\Delta_K\ll \Gamma$), the magnon frequency shift can be amplified by the EP3, which can be probed via the output spectrum of the cavity. The scheme we present provides an alternative approach to measure the MKN in the region $\Delta_K\ll \Gamma$ and has potential applications in designing low-power nonlinear devices based on the MKN.
\end{abstract}

\date{\today}

\maketitle

\section{Introduction}

In the past decade, the progress in cavity-magnon systems has been impressive, where magnons (i.e., collective spin excitations) in ferrimagnetic materials are strongly coupled to photons in microwave cavities via the collective magnetic-dipole interaction~\cite{Lachance-Quirion19,Yuan22,Rameshti22}. Experimentally, the most widely used cavity-magnon system is composed of the millimeter-scale yttrium iron garnet ($\rm{Y}_3\rm{Fe}_5\rm{O}_{12}$ or YIG) crystal and the three-dimensional (3D) microwave cavity~\cite{Tabuchi14,Zhang14,Goryachev14,Zhang15-1}. Up to now, various exotic phenomena have been extensively investigated in cavity-magnon systems, such as magnon dark modes~\cite{Zhang15-Zou}, manipulating spin currents~\cite{Bai15,Bai17}, steady-state magnon-photon entanglement~\cite{Li18}, magnon blockade~\cite{Liu19,Xie20,Wang22-Xiong}, non-Hermitian physics~\cite{Harder17,Cao19,Zhao20}, cooperative polariton dynamics~\cite{Yao17}, enhancing spin-photon coupling~\cite{Hei21}, quantum states of magnons~\cite{Yuan20,Sun21,Zhang22,Qi22}, microwave-to-optical transduction~\cite{Hisatomi16,Zhu20}, and dissipative coupling~\cite{Grigoryan18,Harder18}.

Based on the coherent perfect absorption (CPA), the second-order exceptional point (EP2) was observed~\cite{Zhang17} and the third-order EP (EP3) was subsequently predicted~\cite{You19} in cavity-magnon systems. The CPA refers to a phenomenon that when two (or more) coherent electromagnetic waves are fed into a medium, the waves are completely absorbed by the medium due to both destructive interference between them and medium dissipation, and there are no output waves from the  medium~\cite{Chong10,Wan11}. Intriguing applications of CPA include, e.g., engineering EPs~\cite{Zhang17,You19,Sun14,Wang21}, antilasing~\cite{Wong16,Pichler19}, optical switches~\cite{Fang14,Xiong20-Chen}, and coherent polarization control~\cite{Kang15,Ye16}. The $n$th-order EP (EP$n$) refers to the degenerate point in non-Hermitian systems, where $n$ eigenvalues as well as corresponding $n$ eigenvectors coalesce simultaneously~\cite{Heiss12}. Owing to its fundamental importance and potential applications, the EPs have been explored in various physical systems (see, e.g., Refs.~\cite{Gao15,Lv15,Zhang21-Chen,Jing17,Lu21,Xiong22-Li,Doppler16,Zhang22-Liu,Zhiyenbayev19}). Contrary to the degenerate point in Hermitian systems, the EPs have some unique features. For example, the energy splitting follows a $\epsilon^{1/n}$ dependence around the EP$n$ when the non-Hermitian systems are subjected to a weak perturbation with strength $\epsilon$ ($\ll 1$)~\cite{Wiersig14,Chen17}, which makes it possible to enhance the detection sensitivity~\cite{Liu16,Zhang19-Wang-You,Wang21-Guo}.

It is worth noting that the cavity-magnon system also has reached the nonlinear regime~\cite{Wang16}, where the magnon Kerr nonlinearity (MKN) stems from the magnetocrystalline anisotropy in the YIG~\cite{Zhang-China-19}. The MKN not only results in cavity-magnon bistability~\cite{Wang18,Shen22,Nair21} and tristability~\cite{Shen21,Nair20,Bi21}, nonreciprocal microwave transmission~\cite{Kong19}, and strong long-distance spin-spin coupling~\cite{Xiong22}, but it also leads to magnon-photon  entanglement~\cite{Scully19,Yang21} as well as dynamical quantum phase transition~\cite{Zhang21,Qin22}. In experiments, many phenomena induced by MKN can be detected by measuring the transmission spectrum of the microwave cavity, where the MKN is equivalent to the magnon frequency shift $\Delta_K$ dependent on the magnon population~\cite{Wang16,Zhang-China-19,Wang18,Shen22,Nair21,Shen21,Nair20,Bi21,Kong19}. This probe method works well only when the magnon frequency shift $\Delta_K$ is comparable to (or larger than) the linewidths $\Gamma$ of the peaks in the transmission spectrum of the cavity (i.e., $\Delta_K \geq \Gamma$), while it is not valid in the region $\Delta_K \ll \Gamma$~\cite{Haigh15,Yao17}.

In this paper, we propose a scheme to enhance the detection sensitivity of MKN around an EP in cavity magnonics when $\Delta_K \ll \Gamma$. Here, the considered hybrid system consists of a 3D microwave cavity with two YIG spheres (YIG 1 and YIG 2) embedded (cf.~Fig.~\ref{fig1}), where the magnon mode in YIG 1 has the MKN, while the auxiliary magnon mode in YIG 2 is linear. By feeding two input fields with the same frequency into the 3D microwave cavity via its two ports, an effective pseudo-Hermitian Hamiltonian of the cavity-magnon system can be obtained, where the effective gain of the cavity mode results from the CPA of the two input fields. In the absence of the MKN (corresponding to $\Delta_K=0$), we analyze the eigenvalues of the pseudo-Hermitian Hamiltonian and find the EP3 in the parameter space. Further, we show that the magnon frequency shift $\Delta_K$ ($\ll \Gamma$) induced by the MKN can be amplified by the EP3. Finally, we derive the output spectrum of the 3D cavity and display how the amplification effect can be probed via the output spectrum.

Recently, Ref.~\cite{Nair21-Mukhopadhyay} has proposed to enhance the sensitivity of the magnon-population response to the coefficient of MKN via the anti-parity-time-symmetric phase transition, where the strength of the drive field on the system is fixed. In contrast to Ref.~\cite{Nair21-Mukhopadhyay}, we show that the EP3 can enhance the sensitivity of the eigenvalue response to the small magnon frequency shift induced by MKN in the present work. Our study provides a possibility to detect the MKN in the region $\Delta_K \ll \Gamma$, which is a complement to the existing approach (i.e., measuring the transmission spectrum of the microwave cavity)~\cite{Wang16,Zhang-China-19,Wang18,Shen22,Nair21,Shen21,Nair20,Bi21,Kong19} and may find promising applications in designing low-power nonlinear devices in cavity magnonics. In addition to MKN, other weak signals (such as a weak magnetic field), which can result in the changes of system parameters, can also be detected using our scheme.

\section{The Model}\label{model}

\begin{figure}
\includegraphics[width=0.42\textwidth]{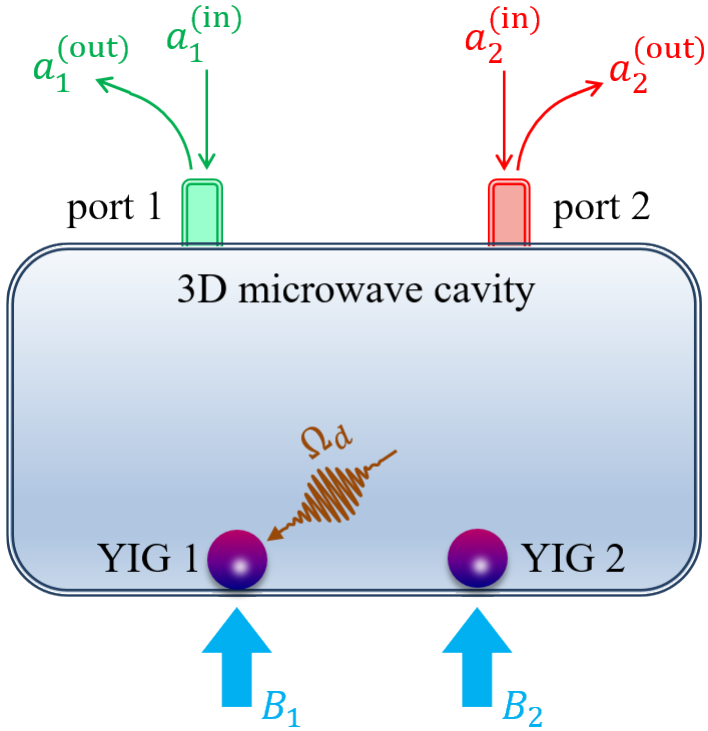}
\caption{Schematic of the proposed setup for enhancing the detection sensitivity of MKN in YIG 1. The cavity magnonic system is composed of two YIG spheres coupled to a 3D microwave cavity, where YIG 1 (YIG 2) is magnetized by a static magnetic field $B_1$ ($B_2$). To measure the weak MKN in YIG 1, one microwave field with Rabi frequency $\Omega_d$ is used to drive YIG 1. In addition, two input fields $a_1^{\rm (in)}$ and $a_2^{\rm (in)}$ are fed into the microwave cavity via ports 1 and 2, respectively, and $a_1^{\rm (out)}$ and $a_2^{\rm (out)}$ denote the corresponding output fields.}
\label{fig1}
\end{figure}

As shown in Fig.~\ref{fig1}, the considered cavity-magnon system consists of two YIG spheres (YIG 1 and YIG 2) and a 3D microwave cavity, where YIG 1 and YIG 2 are uniformly magnetized to saturation by the bias magnetic fields $B_1$ and $B_2$, respectively. Here, to enhance the detection sensitivity of MKN in YIG 1, the YIG 2 provides a magnon mode serving as an ancilla. Now the entire cavity-magnon system is described by the Hamiltonian~\cite{Wang18,Zhang-China-19}
\begin{eqnarray}\label{Hamiltonian}
H&=&\omega_{c}a^{\dag}a+\sum_{j=1,2}\left[\omega_jb_j^{\dag}b_j+K_jb_j^{\dag}b_jb_j^{\dag}b_j+g_j(a^{\dag}b_j+ab_j^{\dag})\right]\nonumber\\
 & &+\Omega_{d}(b_1^{\dag}e^{-i\omega_{\rm{d}}t}+b_1e^{i\omega_{\rm{d}}t}),
\end{eqnarray}
where $a$ and $a^\dag$ ($b_j$ and $b_j^\dag$ with $j=1,2$) are the annihilation and creation operators of the cavity mode (magnon mode in YIG $j$) at frequency $\omega_c$ ($\omega_j$), $g_j$ is the coupling strength between the cavity mode $a$ and the magnon mode $b_j$, and $\Omega_d$ ($\omega_d$) is the strength (frequency) of the drive field on YIG 1. In the two YIG spheres, the magnetocrystalline anisotropy results in the MKN term $K_jb_j^{\dag}b_jb_j^{\dag}b_j$, where the nonlinear coefficient $K_j$ can be continuously tuned from negative values to positive values by adjusting the angle between the crystallographic axis of YIG $j$ and the bias magnetic field $B_j$~\cite{Gurevich96,Stancil09}. Without loss of generality, we assume $K_1>0$ and $K_2=0$ in our scheme. When macroscopic magnons are excited in YIG 1 (i.e., $\langle b_1^\dag b_1\rangle \gg 1$), the system Hamiltonian in Eq.~(\ref{Hamiltonian}) can be linearized as
\begin{eqnarray}\label{linearized-Hamiltonian}
H&=&\omega_{c}a^{\dag}a+\sum_{j=1,2}\left[\omega_jb_j^{\dag}b_j+g_j(a^{\dag}b_j+ab_j^{\dag})\right]+\Delta_K b_1^{\dag}b_1\nonumber\\
 & &+\Omega_{d}(b_1^{\dag}e^{-i\omega_{\rm{d}}t}+b_1e^{i\omega_{\rm{d}}t}),
\end{eqnarray}
with the frequency shift $\Delta_K=2K_1\langle b_1^\dag b_1\rangle$ of the magnon mode $b_1$, where the mean-field approximation $b_1^{\dag}b_1b_1^{\dag}b_1\approx 2\langle b_1^\dag b_1\rangle b_1^\dag b_1$ has been used~\cite{Wang18,Zhang-China-19}.

When the magnon frequency shift $\Delta_K$ is comparable to (or larger than) the linewidths $\Gamma$ of the peaks in the transmission spectrum of the cavity (i.e., $\Delta_K\geq  \Gamma$), the MKN can be probed by measuring the transmission spectrum of the cavity~\cite{Wang16,Zhang-China-19,Wang18,Shen22,Nair21,Shen21,Nair20,Bi21,Kong19}, where the linewidths are comparable to the decay rates of cavity mode and magnon modes. However, in the case of $\Delta_K \ll \Gamma$, it is difficult to probe the MKN in this way~\cite{Haigh15,Yao17}. For measuring the magnon frequency shift $\Delta_K$ in this circumstance, we feed two weak input fields $a_1^{\rm (in)}$ and $a_2^{\rm (in)}$ with same frequency $\omega_p$ into the microwave cavity via ports 1 and 2, respectively. Using the input-output formalism~\cite{Walls94}, we get the equations of motion of the cavity-magnon system as follows:
\begin{eqnarray}\label{Langevin}
\dot{a}&=&-i(\omega_c-i\kappa_c)a
          -\sum_{j=1,2}\left(ig_jb_j-\sqrt{2\kappa_j}\,a_j^{\rm{(in)}}e^{-i\omega_{\rm{p}}t}\right)
                                               +\sqrt{2\kappa_c}\,f_{a}^{\rm(in)},\nonumber\\
\dot{b}_1&=&-i(\omega_1+\Delta_K-i\gamma_1)b_1-ig_1 a-i\Omega_{\rm{d}}e^{-i\omega_{\rm{d}}t}
                                                    +\sqrt{2\gamma_1}\,f_{\rm b1}^{\rm(in)},\nonumber\\
\dot{b}_2&=&-i(\omega_2-i\gamma_2)b_2-ig_2 a +\sqrt{2\gamma_2}\,f_{\rm b2}^{\rm(in)},
\end{eqnarray}
where $\gamma_1$ ($\gamma_2$) is the decay rate of the magnon mode $b_1$ ($b_2$), the total decay rate $\kappa_c=\kappa_{\rm int}+\kappa_1+\kappa_2$ of the cavity mode is composed of the intrinsic decay rate $\kappa_{\rm int}$ and the decay rates $\kappa_1$ and $\kappa_2$ induced by the ports 1 and 2,
and $f_{a}^{\rm(in)}$ ($f_{\rm bj}^{\rm(in)}$) with zero mean value $\langle f_{a}^{\rm(in)}\rangle=0$ ($\langle f_{\rm bj}^{\rm(in)}\rangle=0$) describes the quantum noise from the environment related to the cavity mode (the magnon mode $b_j$). Following the above equations of motion, the expected values $\langle a\rangle$ and $\langle b_j\rangle$ satisfy
\begin{eqnarray}\label{motion}
\langle\dot{a}\rangle&=&-i(\omega_{\rm{c}}-i\kappa_{\rm{c}})\langle a\rangle
          -\sum_{j=1,2}\left(ig_j\langle b_j\rangle-\sqrt{2\kappa_j}\,\langle a_j^{\rm{(in)}}\rangle e^{-i\omega_{\rm{p}}t}\right),\nonumber\\
\langle\dot{b}_1\rangle&=&-i(\omega_1+\Delta_K-i\gamma_1)\langle b_1\rangle-ig_1 \langle a\rangle-i\Omega_{\rm{d}}e^{-i\omega_{\rm{d}}t},\nonumber\\
\langle\dot{b}_2\rangle&=&-i(\omega_2-i\gamma_2)\langle b_2\rangle-ig_2 \langle a\rangle.
\end{eqnarray}
In the absence of the two input fields (corresponding to $\langle a_1^{\rm{(in)}}\rangle=\langle a_2^{\rm{(in)}}\rangle=0$), we denote $\langle a\rangle=\mathcal{A}e^{-i\omega_d t}$ and $\langle b_j\rangle =\mathcal{B}_j e^{-i\omega_d t}$. When the input fields are considered, we assume that the changes of $\langle a\rangle$ and $\langle b_j\rangle$ can be expressed as $Ae^{-i\omega_{\rm{p}}t}$ and $B_je^{-i\omega_{\rm{p}}t}$, i.e.,
\begin{eqnarray}\label{sum}
\langle a\rangle &=&\mathcal{A}e^{-i\omega_{\rm{d}}t}+Ae^{-i\omega_{\rm{p}}t},\nonumber\\
\langle b_j\rangle &=&\mathcal{B}_j e^{-i\omega_{\rm{d}}t}+B_je^{-i\omega_{\rm{p}}t},
\end{eqnarray}
where $|\mathcal{A}| \gg |A|$ and $|\mathcal{B}_j|\gg |B_j|$~\cite{Zhang-China-19}. This assumption is reasonable, because compared with the drive field, the input fields are very weak and can be treated as a perturbation. Now the magnon frequency shift becomes $\Delta_K=2K_1|\mathcal{B}_1|^2$. Substituting Eq.~(\ref{sum}) into Eq.~(\ref{motion}), we have
\begin{eqnarray}\label{mathcal-AB}
\dot{\mathcal{A}}&=&-i(\delta_{\rm cd}-i\kappa_{\rm{c}})\mathcal{A} -ig_{1}\mathcal{B}_1-ig_2\mathcal{B}_2,\nonumber\\
\dot{\mathcal{B}}_1&=&-i(\delta_{\rm 1d}+\Delta_K-i\gamma_1)\mathcal{B}_1-ig_1 \mathcal{A}-i\Omega_{\rm{d}},\nonumber\\
\dot{\mathcal{B}}_2&=&-i(\delta_{\rm 2d}-i\gamma_2)\mathcal{B}_2-ig_2 \mathcal{A},
\end{eqnarray}
and
\begin{eqnarray}\label{AB}
\dot{A}&=&-i(\delta_{\rm cp}-i\kappa_{\rm{c}})A -\sum_{j=1,2}\left(ig_jB_j-\sqrt{2\kappa_j}\,\langle a_j^{\rm{(in)}}\rangle\right),\nonumber\\
\dot{B}_1&=&-i(\delta_{\rm 1p}+\Delta_K-i\gamma_1)B_1-ig_1 A ,\nonumber\\
\dot{B}_2&=&-i(\delta_{\rm 2p}-i\gamma_2)B_2-ig_2 A,
\end{eqnarray}
where $\delta_{\rm cd}=\omega_c-\omega_{d}$ ($\delta_{\rm jd}=\omega_j-\omega_{d}$) is the frequency detuning between the cavity mode (magnon mode $j$) and the drive field, and $\delta_{\rm cp}=\omega_c-\omega_p$ ($\delta_{\rm jp}=\omega_j-\omega_p$) is the frequency detuning between the cavity mode (magnon mode $j$) and the two input fields. Eq.~(\ref{mathcal-AB}) determines the magnon frequency shift $\Delta_K$, while Eq.~(\ref{AB}) determines the output spectrum of the cavity.

According to the input-output theory~\cite{Walls94}, the output field $\langle a_j^{\rm (out)}\rangle$ from the port $j$ of the cavity is given by
\begin{equation}\label{input-output}
\langle a_j^{\rm (out)}\rangle=\sqrt{2\kappa_j}\,A-\langle a_j^{\rm (in)}\rangle.
\end{equation}
Under the pseudo-Hermitian conditions [cf.~Eq.~(\ref{pseudo-Hermitian}) in Sec.~\ref{enhancing}], the CPA may occur by carefully choosing appropriate parameters of the two input fields [cf.~Eqs.~(\ref{strength-CPA}) and (\ref{frequency-CPA}) in Sec.~\ref{enhancing}]~\cite{You19}. The CPA means that the two input fields are nonzero but there are no output fields, i.e., $\langle a_1^{\rm (in)}\rangle \neq 0$ and $\langle a_2^{\rm (in)}\rangle \neq 0$ but $\langle a_1^{\rm (out)}\rangle=\langle a_2^{\rm (out)}\rangle=0$~\cite{Sun14,Wang21,Zhang17}. When $\langle a_1^{\rm (out)}\rangle=\langle a_2^{\rm (out)}\rangle=0$,
\begin{equation}\label{}
\langle a_j^{\rm (in)}\rangle=\sqrt{2\kappa_j}\,A.
\end{equation}
Inserting the above relation into Eq.~(\ref{AB}) to eliminate $\langle a_j^{\rm (in)}\rangle$, Eq.~(\ref{AB}) can be rewritten as
\begin{equation}\label{}
\left(
  \begin{array}{c}
    \dot{A}\\
    \dot{B}_1\\
    \dot{B}_2\\
  \end{array}
\right)=-i H_{\rm eff}\left(
  \begin{array}{c}
      A\\
      B_1\\
      B_2\\
  \end{array}
\right),
\end{equation}
where
\begin{equation}\label{effective}
\begin{split}
H_{\rm eff}=
\left(
  \begin{array}{ccc}
    \delta_{\rm cp}+i\kappa_g & g_{1} & g_{2}\\
    g_{1} & \delta_{\rm 1p}+\Delta_K-i\gamma_{1} & 0\\
    g_{2} & 0 & \delta_{\rm 2p}-i\gamma_{2}\\
  \end{array}
\right)
\end{split}
\end{equation}
is the effective non-Hermitian Hamiltonian of the cavity-magnon system. Due to the occurrence of CPA, the cavity mode has an effective gain $\kappa_g=\kappa_1+\kappa_2-\kappa_{\rm int}$~$(>0)$~\cite{Zhang17,You19}.

\section{Enhancing the detection sensitivity of MKN}\label{enhancing}

\subsection{The EP3 in the cavity-magnon system}

In this section, we study the EP3 in the cavity-magnon system when $\Delta_K=0$. Usually, the eigenvalues of a non-Hermitian Hamiltonian are complex. However, when the system parameters satisfy the pseudo-Hermitian conditions~\cite{You19},
\begin{eqnarray}\label{pseudo-Hermitian}
\kappa_{g}&=&(1+\eta)\gamma_{2},\nonumber\\
\Delta_{2}&=&-\eta \Delta_{1},\nonumber\\
\Delta_{1}^{2}&=&\frac{1+\eta k^{2}}{(1+\eta)\eta}g_{1}^{2}-\gamma_{2}^{2},~~~g_1 \geq g_{\rm min},
\end{eqnarray}
the effective non-Hermitian Hamiltonian $H_{\rm eff}$ in Eq.~(\ref{effective}) has the pseudo-Hermiticity and thus can also own either three real eigenvalues or one real and two complex-conjugate eigenvalues~\cite{Mostafazadeh021,Mostafazadeh022,Mostafazadeh023}. The parameter $\eta=\gamma_{1}/\gamma_{2}$ ($k=g_{2}/g_{1}$) denotes the ratio between the decay rates $\gamma_1$ and $\gamma_2$ (coupling strengths $g_1$ and $g_2$), $\Delta_{j}=\omega_{j}-\omega_c$ is the frequency detuning of the magnon mode~$j$ relative to the cavity mode, and $g_{\rm min}=[(1+\eta)\eta/(1+\eta k^{2})]^{1/2} \gamma_{2}$ is the allowed minimal value of the coupling strength $g_1$ for ensuring $\Delta_{1}^{2} \geq 0$.

For engineering the EP3 under the pseudo-Hermitian conditions in Eq.~(\ref{pseudo-Hermitian}), the parameters $\eta$ and $k$ must satisfy  the following constraint~\cite{You19}:
\begin{equation}\label{eta-k}
k      =\left(\frac{1+2\eta}{2\eta+\eta^2}\right)^{3/2}.
\end{equation}
In the symmetric case of $\eta=k=1$, the non-Hermitian Hamiltonian $H_{\rm eff}$ has three eigenvalues, $\Omega_0=\delta_{\rm cp}$ and $\Omega_\pm=\delta_{\rm cp}\pm \sqrt{3g_1^2-4\gamma_2^2}$~\cite{You19}. Obviously, $\Omega_0$ is real and independent of the coupling strength $g_1$ and the decay rate $\gamma_2$, while $\Omega_\pm$ are functions of $g_1$ and $\gamma_2$. To have three real eigenvalues, the coupling strength $g_1$ should be in the region $g_1>g_{\rm EP3}$, where $g_{\rm EP3}=2\gamma_2/\sqrt{3}$. For $g_1=g_{\rm EP3}$ in particular, the three eigenvalues $\Omega_\pm$ and $\Omega_0$ coalesce to $\Omega_\pm=\Omega_0=\Omega_{\rm EP3}=\delta_{\rm cp}$, and the corresponding three eigenvectors of $H_{\rm eff}$ also coalesce to $|\alpha\rangle_\pm=|\alpha\rangle_0=|\alpha\rangle_{\rm EP3}=\frac{1}{\sqrt{3}}\left(1,-\frac{1+\sqrt{3}i}{2},\frac{1-\sqrt{3}i}{2}\right)^{T}$. This coalescent point at $g_1=g_{\rm EP3}$ is referred to as the EP3. While $g_{\rm min} \leq g_1 <g_{\rm EP3}$, $\Omega_\pm$ become complex. For the asymmetric case with $\eta \neq 1$ and $k \neq 1$, the expressions of $\Omega_\pm$ and $\Omega_0$ are cumbersome and not shown here, and we only give the coalesced eigenvalues $\Omega_\pm=\Omega_0=\Omega_{\rm EP3}$ at $g_1=g_{\rm{EP3}}=[2\eta(\eta^2+2\eta)^{1/2}/(1+2\eta)]\gamma_{2}$, where~\cite{You19}
\begin{equation}\label{EP3-eigenvalues}
\Omega_{\rm EP3}=\delta_{\rm cp}-\frac{\sqrt{3}(\eta-1)\eta}{2\eta^2+5\eta+2}\gamma_2.
\end{equation}
At the EP3, the three eigenvectors of $H_{\rm eff}$ coalesce to
\begin{equation}\label{EP3-eigenvectors}
|\alpha\rangle_{\rm EP3}=\frac{1}{\sqrt{\mathcal{N}}}
       \left(1,-\frac{2\sqrt{\eta^2+2\eta}}{\sqrt{3}-i(1+2\eta)},\frac{2\sqrt{2\eta+1}}{\sqrt{3}\eta+i(2+\eta)}\right)^{T},
\end{equation}
with the normalization factor $\mathcal{N}=(2\eta^2+5\eta+2)/(\eta^2+\eta+1)$, i.e., $|\alpha\rangle_\pm=|\alpha\rangle_0=|\alpha\rangle_{\rm EP3}$. Note that the results in Eqs.~(\ref{EP3-eigenvalues}) and (\ref{EP3-eigenvectors}) are also valid for the symmetric case of $\eta=k=1$.

As stated in Sec.~\ref{model}, the effective non-Hermitian Hamiltonian $H_{\rm eff}$ in Eq.~(\ref{effective}) is obtained in the presence of CPA. For engineering the CPA in the pseudo-Hermitian conditions in Eq.~(\ref{pseudo-Hermitian}), the strengths of the two input fields should satisfy~\cite{You19}
\begin{equation}\label{strength-CPA}
\frac{\langle a_2^{\rm (in)}\rangle}{\langle a_1^{\rm (in)}\rangle}=\sqrt{\frac{\kappa_2}{\kappa_1}}.
\end{equation}
In addition, the same frequency of the two input fields need to be equal to the real eigenvalues of $H_{\rm eff}$~\cite{You19}, i.e.,
\begin{equation}\label{frequency-CPA}
\omega_p^{\rm (CPA)}=\Omega_{\pm,0}~~{\rm when\,\, Im}[\Omega_{\pm,0}]=0.
\end{equation}
Therefore, the eigenvalues and the EP3 of the pseudo-Hermitian cavity-magnon system can be probed by measuring the CPA via the output spectrum of the cavity in experiments~\cite{Sun14,Zhang17,Wang21}.

\begin{figure}
\includegraphics[width=0.48\textwidth]{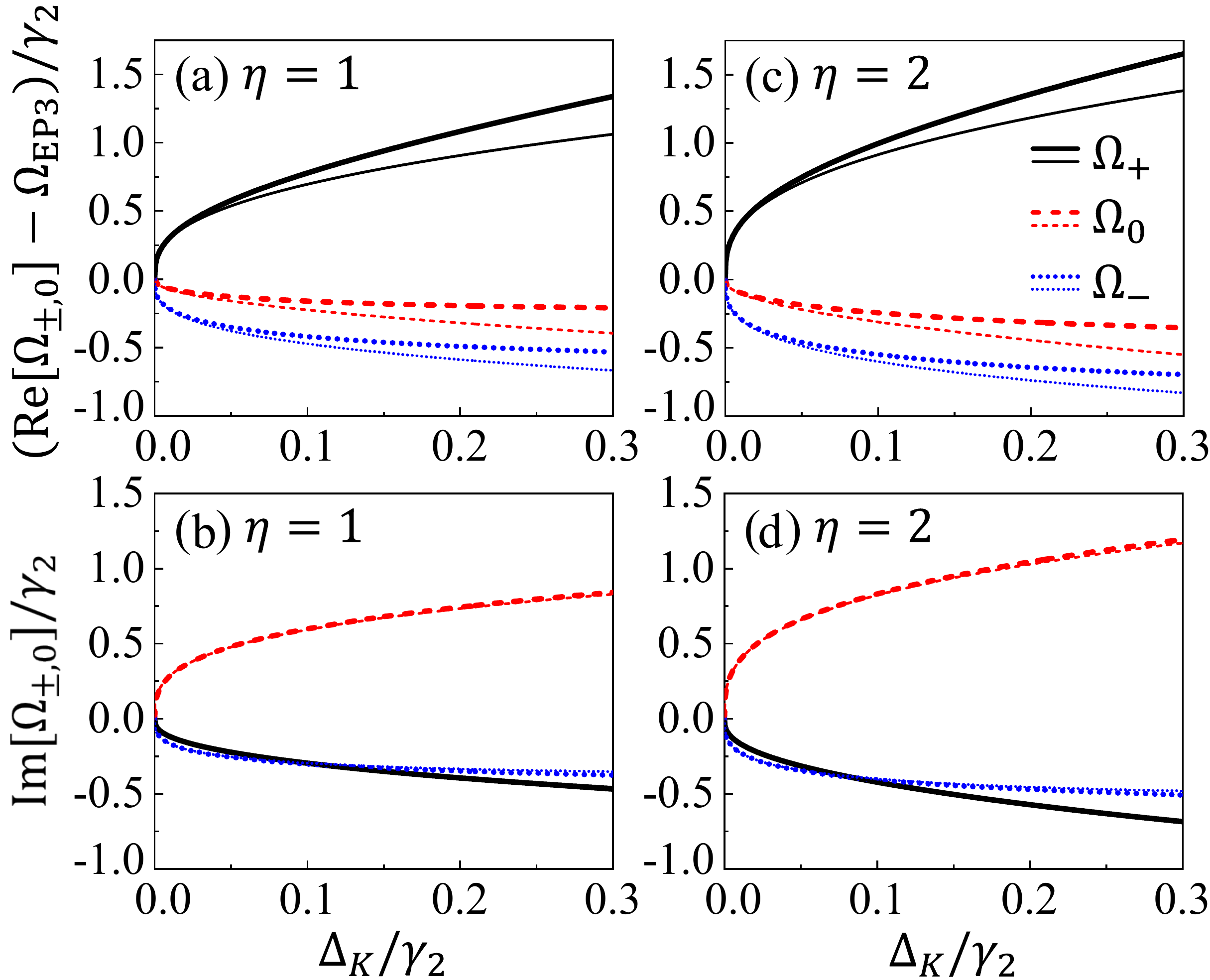}
\caption{The changes of the real and imaginary parts of the eigenvalues $\Omega_{\pm}$ and $\Omega_0$, $({\rm Re}[\Omega_{\pm,0}]-\Omega_{\rm EP3})/\gamma_2$ and ${\rm Im}[\Omega_{\pm,0}]/\gamma_2$, versus the magnon frequency shift $\Delta_K/\gamma_2$ near the EP3, where $\eta=1$ in (a,b), while $\eta=2$ in (c,d). In (a)--(d), the thick curves correspond to the numerical results obtained by numerically solving the characteristic equation in Eq.~(\ref{characteristic}), and the thin curves correspond to the analytical results in Eq.~(\ref{eigenvalues-EP3}). Note that the thick curves almost overlap the thin curves in (b,d).}
\label{fig2}
\end{figure}

\subsection{Eigenvalue response to the MKN near the EP3}

Here we investigate the eigenvalue response to the MKN in YIG 1 near the EP3. Considering the magnon frequency shift $\Delta_K$ ($\neq 0$), the three eigenvalues of the cavity-magnon system can be obtained by solving the corresponding characteristic equation
\begin{equation}\label{characteristic}
|H_{\rm eff}-\Omega I|=0,
\end{equation}
with an identity matrix $I$. Because the magnon frequency shift $\Delta_K$ is much smaller than other parameters of the cavity-magnon system, we can perturbatively expand the eigenvalue $\Omega$ near the EP3 as
\begin{equation}\label{Omega}
\Omega = \Omega_{\rm EP3}+\lambda_1\xi^{1/3}\gamma_2+\lambda_2\xi^{2/3}\gamma_2
\end{equation}
using a Newton-Puiseux series~\cite{Hodaei17,Xiong21,Zeng21}, where only the first two terms are considered, and $\Omega_{\rm EP3}$ is given in Eq.~(\ref{EP3-eigenvalues}). The coefficients $\lambda_1$ and $\lambda_2$ are complex, while $\xi=\Delta_K/\gamma_2$ $( \ll 1)$ is real. With Eq.~(\ref{Omega}), the characteristic equation of the cavity-magnon system in Eq.~(\ref{characteristic}) can be expressed as
\begin{equation}\label{f1xi}
f_{1}\xi+f_{4/3}\xi^{4/3}+f_{5/3}\xi^{5/3}+f_{2}\xi^{2}+f_{7/3}\xi^{7/3}=0,
\end{equation}
where the coefficients are
\begin{eqnarray}\label{}
f_{1}&=&\lambda_1^3-\frac{4\eta^2(1-\sqrt{3}i)}{1+2\eta},\nonumber\\
f_{4/3}&=&3\lambda_1^2\lambda_2-\frac{2\eta[\sqrt{3}-i(1+2\eta)]}{1+2\eta}\lambda_1,\nonumber\\
f_{5/3}&=&3\lambda_1\lambda_2^2-2\lambda_1^2-\frac{2\eta[\sqrt{3}-i(1+2\eta)]}{1+2\eta}\lambda_2,\nonumber\\
f_{2}&=&\lambda_2^3-4\lambda_1\lambda_2 ,\nonumber\\
f_{7/3}&=&-2\lambda_2^2.
\end{eqnarray}
Since $\xi \gg \xi^{4/3}\gg \xi^{5/3}\gg \xi^{2}\gg \xi^{7/3}$, we can ignore the contributions from the last three terms in Eq.~(\ref{f1xi}), and Eq.~(\ref{f1xi}) is reduced to $f_{1}\xi+f_{4/3}\xi^{4/3}=0$. To ensure the relation $f_{1}\xi+f_{4/3}\xi^{4/3}=0$ is valid for any $\xi$, the coefficients $f_{1}$ and $f_{4/3}$ must be zero, i.e., $f_{1}=f_{4/3}=0$. Solving $f_{1}=f_{4/3}=0$, we obtain three sets of solutions for the coefficients $\lambda_1$ and $\lambda_2$,
\begin{eqnarray}\label{}
\lambda_1^{(l)}&=&\left(\frac{8\eta^2}{1+2\eta}\right)^{1/3}e^{i\theta_l},\nonumber\\
\lambda_2^{(l)}&=&\frac{2\eta[\sqrt{3}-i(1+2\eta)]}{3(1+2\eta)\lambda_1^{(l)}},
\end{eqnarray}
with $l=\pm,0$, where $\theta_+=17\pi/9$, $\theta_-=11\pi/9$, and $\theta_0=5\pi/9$. Now the three complex eigenvalues of the cavity-magnon system read
\begin{eqnarray}\label{eigenvalues-EP3}
\Omega_{+}& = &\Omega_{\rm EP3}+\lambda_1^{(+)}\xi^{1/3}\gamma_2+\lambda_2^{(+)}\xi^{2/3}\gamma_2,\nonumber\\
\Omega_{0}& = &\Omega_{\rm EP3}+\lambda_1^{(0)}\xi^{1/3}\gamma_2+\lambda_2^{(0)}\xi^{2/3}\gamma_2,\nonumber\\
\Omega_{-}& = &\Omega_{\rm EP3}+\lambda_1^{(-)}\xi^{1/3}\gamma_2+\lambda_2^{(-)}\xi^{2/3}\gamma_2.
\end{eqnarray}
Clearly, the changes of the eigenvalues, $\Omega_{\pm,0}-\Omega_{\rm EP3}$, are proportional to $\xi^{1/3}$ in the case of $\xi\ll 1$, i.e., $\Omega_{\pm,0}-\Omega_{\rm EP3} \approx \lambda_1^{(\pm,0)}\xi^{1/3}\gamma_2$.

By numerically solving the characteristic equation in Eq.~(\ref{characteristic}), we further study the eigenvalue response to the MKN near the EP3 when $\Delta_K/\gamma_2<0.3$. In the symmetric case of $\eta=1$, we plot the changes of the real and imaginary parts of $\Omega_{\pm}$ and $\Omega_0$, $({\rm Re}[\Omega_{\pm,0}]-\Omega_{\rm EP3})/\gamma_2$ and ${\rm Im}[\Omega_{\pm,0}]/\gamma_2$, as functions of magnon frequency shift $\Delta_K/\gamma_2$ (i.e., $\xi$) in Figs.~\ref{fig2}(a) and \ref{fig2}(b), where the thick curves correspond to the numerical results, and the thin curves correspond to the analytical results in Eq.~(\ref{eigenvalues-EP3}). The analytical results and the numerical results are almost consistent for $\Delta_K/\gamma_2<0.1$, while the analytical results deviate from the numerical results when $\Delta_K/\gamma_2>0.1$ because the condition $\Delta_K/\gamma_2 \ll 1$ has been used in deriving Eq.~(\ref{eigenvalues-EP3}). Obviously, $({\rm Re}[\Omega_{\pm,0}]-\Omega_{\rm EP3})/\gamma_2$ and ${\rm Im}[\Omega_{\pm,0}]/\gamma_2$ versus $\Delta_K/\gamma_2$ sharply change. This is because the small frequency shift $\Delta_K$ is amplified by the EP3~\cite{Wiersig14,Chen17}. In the region $\xi \ll 1$, $({\rm Re}[\Omega_{\pm,0}]-\Omega_{\rm EP3})/\gamma_2$ and ${\rm Im}[\Omega_{\pm,0}]/\gamma_2$ follow the cube-root of $\xi$, i.e., $({\rm Re}[\Omega_{\pm,0}]-\Omega_{\rm EP3})/\gamma_2 \approx {\rm Re}[\lambda_1^{(\pm,0)}] \xi^{1/3}$ and ${\rm Im}[\Omega_{\pm,0}]/\gamma_2 \approx {\rm Im}[\lambda_1^{(\pm,0)}] \xi^{1/3}$. It is very different from the existing approach of measuring MKN, where the energy splitting follows a $\xi$ dependence~\cite{Wang16,Zhang-China-19,Wang18}. Further, we find that the amplification effect is more significant for a larger value of $\eta$ [cf.~Figs.~\ref{fig2}(a) and \ref{fig2}(c); Figs.~\ref{fig2}(b) and \ref{fig2}(d)], which results from the monotonous increase of $|\lambda_1^{(l)}|=[8\eta^2/(1+2\eta)]^{1/3}$ versus $\eta$. Considering the experimentally accessible parameters, we choose $1 \leq \eta \leq 3$ in our study~\cite{Lachance-Quirion19,Zhang17,You19}. This amplification effect of the EP3 can be used to measure the MKN in the case of $\Delta_K/\gamma_2 < 1$ (cf. Sec.~\ref{readout}).

\section{Measuring the MKN via the output spectrum of the cavity}\label{readout}

\begin{figure}
\includegraphics[width=0.45\textwidth]{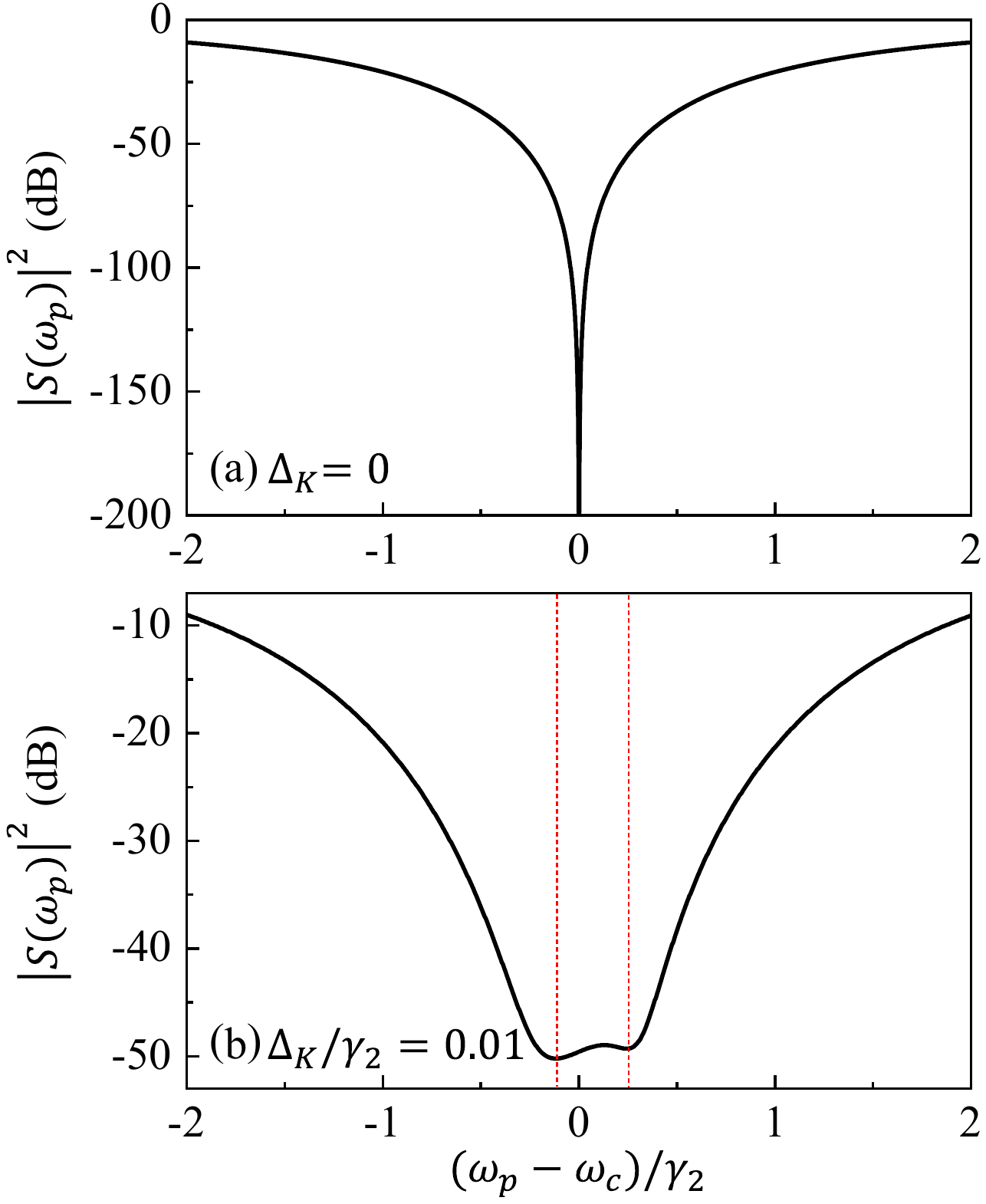}
\caption{(a) The output spectrum $|S(\omega_p)|^2$ of the cavity at the EP3, where $\Delta_K=0$. (b) The output spectrum $|S(\omega_p)|^2$ of the cavity near the EP3 when $\Delta_K \neq 0$ (e.g., $\Delta_K/\gamma_2=0.01$). The (red) dashed vertical lines in (b) highlight the locations of the two dips in the output spectrum. Other parameters are chosen to be $\gamma_1/\gamma_2=1$, $\kappa_{\rm int}/\gamma_2=1$, and $\kappa_1/\gamma_2=\kappa_2/\gamma_2=1.5$.}
\label{fig3}
\end{figure}

\begin{figure}
\includegraphics[width=0.43\textwidth]{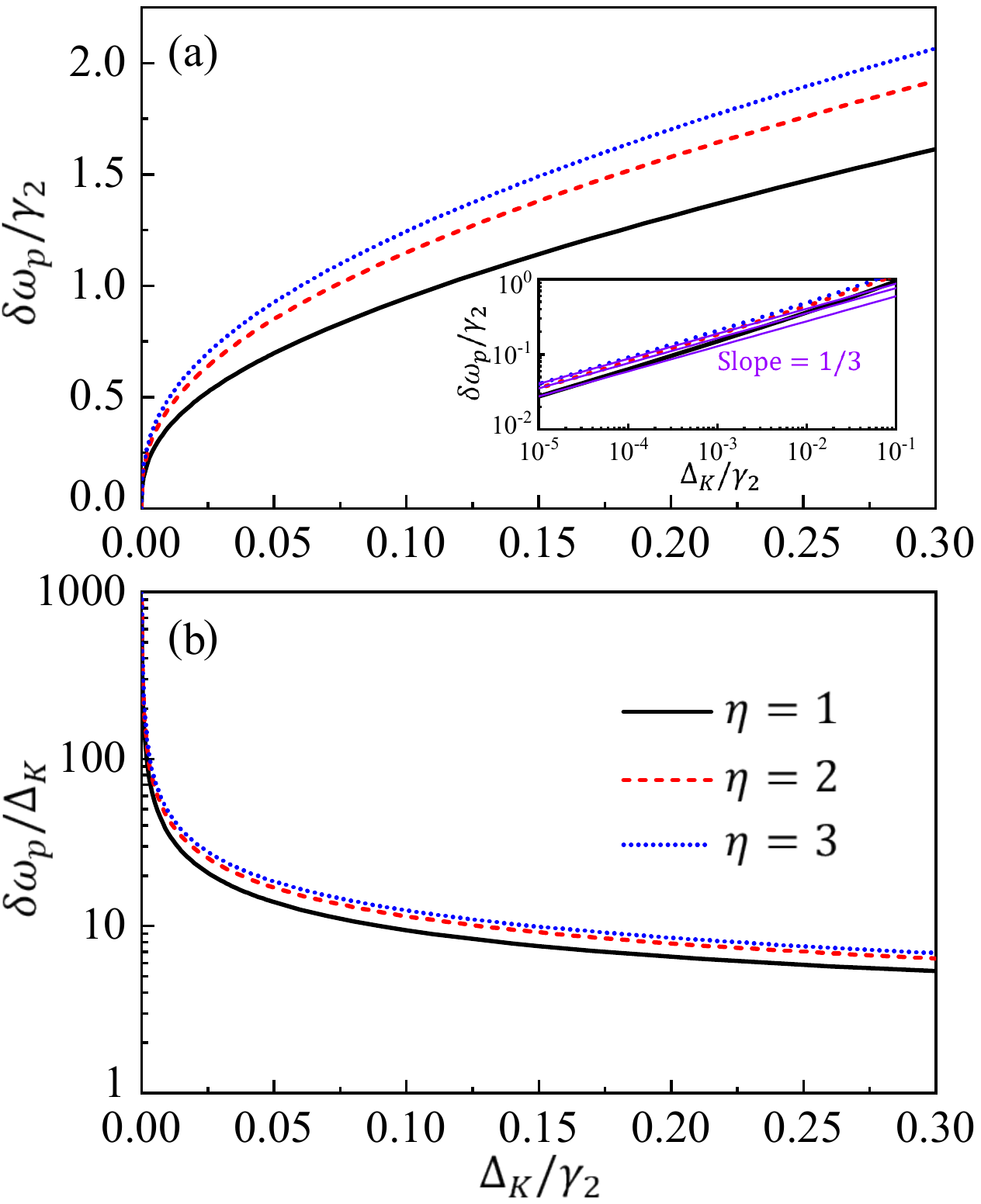}
\caption{(a) The distance $\delta\omega_p/\gamma_2$ between the two dips in the output spectrum of the cavity versus the magnon frequency shift $\Delta_K/\gamma_2$ for different $\eta$. The inset displays the logarithmic relationship between $\delta\omega_p/\gamma_2$ and $\Delta_K/\gamma_2$ for different $\eta$, where the three (violet) thin curves with a same slope of $1/3$ serve as guides to the eyes. (b) Detection sensitivity enhancement factor $\delta\omega_p/\Delta_K$ versus the magnon frequency shift $\Delta_K/\gamma_2$ for different $\eta$. Here $\eta=1$ for the (black) solid curve, $\eta=2$ for the (red) dashed curve, and $\eta=3$ for the (blue) dotted curve. Other parameters are chosen to be $\gamma_1/\gamma_2=\eta$, $\kappa_{\rm int}/\gamma_2=1$, and $\kappa_1/\gamma_2=\kappa_2/\gamma_2=1+0.5\eta$.}
\label{fig4}
\end{figure}

In the cavity-magnon system, we can measure the eigenvalue response to the MKN via the output spectrum of the cavity~\cite{Zhang17,You19}. In the theory, the output spectrum can be derived using Eqs.~(\ref{AB}) and (\ref{input-output}). At the steady state, we solve Eq.~(\ref{AB}) with $\dot{A}=\dot{B}_1=\dot{B}_2=0$ and obtain the change $A$ of the cavity field $\langle a\rangle$ due to the two input fields,
\begin{equation}\label{}
A=\frac{\sqrt{2\kappa_{1}}\langle a_{1}^{\rm{(in)}}\rangle+\sqrt{2\kappa_{2}}\langle a_2^{\rm{(in)}}\rangle}{\kappa_c+i\delta_{\rm cp}+\sum(\omega_p)},
\end{equation}
where
\begin{equation}\label{}
\sum(\omega_p)=\frac{g_1^2}{\gamma_1+i(\delta_{\rm 1p}+\Delta_K)}+\frac{g_2^2}{\gamma_2+i\delta_{\rm 2p}}
\end{equation}
is the self-energy. Correspondingly, the two output fields $\langle a_1^{\rm (out)}\rangle$ and $\langle a_2^{\rm (out)}\rangle$ in Eq.~(\ref{input-output}) can be expressed as
\begin{eqnarray}\label{Hamiltonian-drive}
\langle a_1^{\rm (out)}\rangle&=&\frac{2\kappa_{1}\langle a_{1}^{\rm{(in)}}\rangle +2\sqrt{\kappa_{1}\kappa_{2}}\langle a_2^{\rm{(in)}}\rangle}
                                {\kappa_c+i\delta_{\rm cp}+\sum(\omega_p)}-\langle a_1^{\rm (in)}\rangle,\nonumber\\
\langle a_2^{\rm (out)}\rangle&=&\frac{2\sqrt{\kappa_{1}\kappa_{2}}\langle a_{1}^{\rm{(in)}}\rangle+2\kappa_{2}\langle a_2^{\rm{(in)}}\rangle}
                                {\kappa_c+i\delta_{\rm cp}+\sum(\omega_p)}-\langle a_2^{\rm (in)}\rangle.
\end{eqnarray}
It follows from Eq.~(\ref{Hamiltonian-drive}) that $\langle a_1^{\rm (out)}\rangle=S(\omega_p) \langle a_1^{\rm (in)}\rangle$ and $\langle a_2^{\rm (out)}\rangle=S(\omega_p) \langle a_2^{\rm (in)}\rangle$ under the constraint in Eq.~(\ref{strength-CPA}), where
\begin{equation}\label{output-spectrum}
S(\omega_p)=\frac{2\kappa_{1}+2\kappa_{2}}{\kappa_c+i\delta_{\rm cp}+\sum(\omega_p)}-1
\end{equation}
is the output spectrum of the microwave cavity. It can be easily verified that in the case of $\Delta_K=0$, the output spectrum $S(\omega_p)$ is zero [i.e., $S(\omega_p)=0$] when the system parameters satisfy the pseudo-Hermitian conditions in Eq.~(\ref{pseudo-Hermitian}) and the same frequency of the two input fields is given in Eq.~(\ref{frequency-CPA})~\cite{You19}.

At the EP3, the three eigenvalues $\Omega_\pm$ and $\Omega_0$ of the cavity-magnon system coalesce to $\Omega_{\rm EP3}$, and the CPA occurs at $\omega_p^{\rm (CPA)}=\Omega_{\rm EP3}$, i.e., there is only one CPA point with $|S(\omega_p)|=0$ in the output spectrum [see Fig.~\ref{fig3}(a)]. In the presence of the MKN (i.e., $\Delta_K \neq 0$), the CPA disappears, and there are two dips in the output spectrum highlighted by the two (red) dashed vertical lines in Fig.~\ref{fig3}(b). The locations and linewidths of the dips in the output spectrum are determined by the real and imaginary parts of the complex eigenvalues of the cavity-magnon system given in Eq.~(\ref{eigenvalues-EP3}). The left dip at $\omega_p^{\rm (dip1)} \approx {\rm Re}[\Omega_-]$ (right dip at $\omega_p^{\rm (dip2)} \approx {\rm Re}[\Omega_+]$) corresponds to the eigenvalue $\Omega_-$ ($\Omega_+$). Note that because $|{\rm Im}[\Omega_0]|>|{\rm Im}[\Omega_\pm]|$ [cf.~Figs.~\ref{fig2}(b) and \ref{fig2}(d)], there is no dip in the output spectrum corresponding to the eigenvalue $\Omega_0$. Therefore, we can measure the MKN by the output spectrum of the cavity.

To characterize the detection sensitivity enhancement of MKN near the EP3, we introduce an experimentally measurable quantity
\begin{equation}\label{}
\delta\omega_p=\omega_p^{\rm (dip2)}-\omega_p^{\rm (dip1)},
\end{equation}
which presents the distance between the two dips in the output spectrum of the cavity. By numerically solving the output spectrum $S(\omega_p)$ in Eq.~(\ref{output-spectrum}), we plot the frequency difference $\delta\omega_p/\gamma_2$ as a function of the magnon frequency shift $\Delta_K/\gamma_2$ for different values of $\eta$ in Fig.~\ref{fig4}(a), where $\delta\omega_p/\gamma_2$ increases monotonically with $\Delta_K/\gamma_2$. Obviously, for a given value of $\Delta_K/\gamma_2$, the corresponding frequency difference $\delta\omega_p/\gamma_2$ between the two dips is far larger than the magnon frequency shift $\Delta_K/\gamma_2$, i.e., $\delta\omega_p \gg \Delta_K$.  In contrast, the frequency difference induced by $\Delta_K$ is approximately equal to $\Delta_K$ in the existing approach of measuring MKN~\cite{Wang16,Zhang-China-19,Wang18}. This means that the magnon frequency shift $\Delta_K$ is amplified by the EP3. For sufficiently small $\Delta_K/\gamma_2$, $\delta\omega_p$ follows a $(\Delta_K/\gamma_2)^{1/3}$ dependence [see the inset in Fig.~\ref{fig4}(a)]. Especially, for a larger value of $\eta$, the amplification effect of the EP3 is more significant. Moreover, we also display the detection sensitivity enhancement factor $\delta\omega_p/\Delta_K$ versus the magnon frequency shift $\Delta_K/\gamma_2$ in Fig.~\ref{fig4}(b), where $\delta\omega_p/\Delta_K$ monotonically decreases for different $\eta$. In the region $\Delta_K/\gamma_2 \ll 1$, $\delta\omega_p/\Delta_K$ is proportional to $(\Delta_K/\gamma_2)^{-2/3}$. When $\Delta_K/\gamma_2$ tends to $0$, the sensitivity enhancement factor $\delta\omega_p/\Delta_K$ tends to infinity, i.e., $\delta\omega_p/\Delta_K$ diverges at $\Delta_K/\gamma_2=0$.

\section{Discussions and conclusions}

In our study, all results are based on the equations of motion in Eq.~(\ref{motion}), which describes the average behavior of the cavity-magnon system  in the mean-field approximation by neglecting the impacts of noises [including classical noise related to fluctuations of system parameters and quantum noise related to terms $\sqrt{2\kappa_c}\,f_{a}^{\rm(in)}$ and $\sqrt{2\gamma_j}\,f_{\rm bj}^{\rm(in)}$ in Eq.~(\ref{Langevin})] and quantum fluctuations (related to $\delta a=a-\langle a\rangle$ and $\delta b_j=b_j-\langle b_j\rangle$). Using Eq.~(\ref{motion}), we investigate the detection sensitivity enhancement of MKN by deriving the effective non-Hermitian Hamiltonian $H_{\rm eff}$ of the cavity-magnon system in Eq.~(\ref{effective}) and the output spectrum $S(\omega_p)$ of the microwave cavity in Eq.~(\ref{output-spectrum}). This procedure is widely applied in studying EP-based sensors~\cite{Wiersig14,Chen17,Liu16,Zhang19-Wang-You,Wang21-Guo}, and the related theoretical predictions have been demonstrated experimentally in various physical systems~\cite{Wiersig20}. For example, the detection sensitivity enhancement factor of 23 has been realized experimentally in a ternary micro-ring system~\cite{Hodaei17}.

However, in the region with the signal being comparable to the noises and quantum fluctuations, the impacts of noises and quantum fluctuations on the EP-based sensor should be considered~\cite{Wiersig20}. The classical noise caused by the technical limitation can reduce the resolvability of frequency difference $\delta\omega_p$ by broadening the linewidth of the output spectrum $S(\omega_p)$~\cite{Wolff19,Wiersig20-PRA}. In principle, the classical noise can be made arbitrarily small in the cavity-magnon system. Different from the classical noise, the quantum noise cannot be made arbitrarily small owing to the vacuum noise. Due to the quantum noise and quantum fluctuations, the diverging sensitivity enhancement factor [cf. Fig.~\ref{fig4}(b) and related discussions] does not necessarily lead to arbitrary high measurement precision, where the measurement precision refers to the smallest measurable change of signal~\cite{Langbein18,Lau18,Chen-Jin19,Zhang-Sweeney19}. This is because the EP-based sensor is sensitive to not only the signal but also the quantum noise, and thus the quantum-limited signal-to-noise ratio cannot be improved~\cite{Wiersig20}. Following the procedures in Refs.~\cite{Lau18,Chen-Jin19,Zhang-Sweeney19}, one can derive the upper bound of the signal-to-noise ratio by calculating the quantum Fisher information based on Heisenberg-Langevin equations in Eq.~(\ref{Langevin}). For the MKN term $K_1b_1^\dag b_1b_1^\dag b_1$, the corresponding effective Hamiltonian for quantum fluctuations can be expressed as $H_{\rm flu}=2\Delta_K\delta b_1^\dag \delta b_1 + \,\chi\delta b_1^\dag\delta b_1^\dag+\chi^*\delta b_1\delta b_1$ with $\chi=K_1\langle b_1\rangle^2$~\cite{Xiong22,Scully19,Yang21}. The two-magnon terms $\,\chi\delta b_1^\dag\delta b_1^\dag$ and $\chi^*\delta b_1\delta b_1$ can squeeze the quantum fluctuations of magnon mode $b_1$, which can be transferred to cavity mode $a$ and magnon mode $b_2$ via their interactions and leads to the squeezing of cavity mode $a$ and magnon mode $b_2$~\cite{Yang21}. The squeezing of quantum fluctuations induced by MKN may be helpful for improving the measurement precision~\cite{Kruse16,Malnou19}.

Before concluding, we briefly analyze the experimental feasibility of the present scheme. In cavity magnonics, both the intrinsic decay rate of the 3D microwave cavity as well as the decay rate of the magnon mode are of the order 1~MHz (i.e., $\kappa_{\rm int}/2\pi \sim $1~MHz and $\gamma_{1,2}/2\pi \sim $1~MHz)~\cite{Lachance-Quirion19}, while the decay rates $\kappa_{1,2}$ due to the two ports of the cavity can be tuned from 0 to 8~MHz~\cite{Zhang17}. Since the frequency of the magnon mode in the YIG is proportional to the bias magnetic field, the frequencies $\omega_{1,2}$ can be easily controlled~\cite{Zhang15-Zou,Shen21}. In Ref.~\cite{Zhang17}, the EP2 based on CPA has been observed, where the cavity-magnon coupling can be adjusted (ranging from 0 to 9~MHz) via moving the YIG sphere, and the relative amplitudes (relative phases) of the two input fields, $\langle a_1^{\rm (in)}\rangle$ and $\langle a_2^{\rm (in)}\rangle$, are also tunable via a variable attenuator (a phase shifter). In addition, the magnon frequency shift $\Delta_K$ caused by the MKN is dependent on the strength of the drive field on the magnon mode~\cite{Wang16,Wang18,Shen22}. These available conditions ensure that our scheme in the present work is experimentally accessible.

In conclusion, we have presented a feasible scheme to enhance the detection sensitivity of MKN via the CPA around an EP3. In the proposed scheme, the cavity-magnon system consists of a 3D microwave cavity and two YIG spheres. With the assistance of the CPA, an effective pseudo-Hermitian Hamiltonian of the cavity-magnon system can be obtained, which makes it possible to engineer the EP3 in the parameter space. Considering the magnon frequency shift caused by the MKN, we find that it can be amplified by the EP3. Moreover, we show that this amplification effect can be measured using the output spectrum of the 3D cavity. Our proposal paves a way to measure the MKN in the case of $\Delta_K \ll \Gamma$.

\section*{Acknowledgments}

This work is supported by the National Natural Science Foundation of China (Grant No. 12205069) and the key program of the Natural Science Foundation of Anhui (Grant No. KJ2021A1301).


\begin{thebibliography}{99}

\bibitem{Lachance-Quirion19}
D. Lachance-Quirion, Y. Tabuchi, A. Gloppe, K. Usami, and Y. Nakamura,
Hybrid quantum systems based on magnonics,
Appl. Phys. Express \textbf{12}, 070101 (2019).

\bibitem{Yuan22}
H. Y. Yuan, Y. Cao, A. Kamra, R. A. Duine, and P. Yan,
Quantum magnonics: When magnon spintronics meets quantum information science,
Phys. Rep. \textbf{965}, 1 (2022).

\bibitem{Rameshti22}
B. Z. Rameshti, S. V. Kusminskiy, J. A. Haigh, K. Usami, D. Lachance-Quirion, Y. Nakamura, C. M. Hu, H. X. Tang, G. E. W. Bauer, and Y. M. Blanter,
Cavity magnonics,
Phys. Rep. \textbf{979}, 1 (2022).

\bibitem{Tabuchi14}
Y. Tabuchi, S. Ishino, T. Ishikawa, R. Yamazaki, K. Usami, and Y. Nakamura,
Hybridizing Ferromagnetic Magnons and Microwave Photons in the Quantum Limit,
Phys. Rev. Lett. \textbf{113}, 083603 (2014).

\bibitem{Zhang14}
X. Zhang, C. L. Zou, L. Jiang, and H. X. Tang,
Strongly Coupled Magnons and Cavity Microwave Photons,
Phys. Rev. Lett. \textbf{113}, 156401 (2014).

\bibitem{Goryachev14}
M. Goryachev, W. G. Farr, D. L. Creedon, Y. Fan, M. Kostylev, and M. E. Tobar,
High-Cooperativity Cavity QED with Magnons at Microwave Frequencies,
Phys. Rev. Applied \textbf{2}, 054002 (2014).

\bibitem{Zhang15-1}
D. Zhang, X. M. Wang, T. F. Li, X. Q. Luo, W. Wu, F. Nori, and J. Q. You,
Cavity quantum electrodynamics with ferromagnetic magnons in a small yttrium-iron-garnet sphere,
npj Quantum Inf. \textbf{1}, 15014 (2015).

\bibitem{Zhang15-Zou}
X. Zhang, C. L. Zou, N. Zhu, F. Marquardt, L. Jiang, and H. X. Tang,
Magnon dark modes and gradient memory,
Nat. Commun. \textbf{6}, 8914 (2015)

\bibitem{Bai15}
L. Bai, M. Harder, Y. P. Chen, X. Fan, J. Q. Xiao, and C. M. Hu,
Spin Pumping in Electrodynamically Coupled Magnon-Photon Systems,
Phys. Rev. Lett. \textbf{114}, 227201 (2015).

\bibitem{Bai17}
L. Bai, M. Harder, P. Hyde, Z. Zhang, C. M. Hu, Y. P. Chen, and J. Q. Xiao,
Cavity Mediated Manipulation of Distant Spin Currents Using a Cavity-Magnon-Polariton,
Phys. Rev. Lett. \textbf{118}, 217201 (2017).

\bibitem{Li18}
J. Li, S. Y. Zhu, and G. S. Agarwal,
Magnon-Photon-Phonon Entanglement in Cavity Magnomechanics,
Phys. Rev. Lett. \textbf{121}, 203601 (2018).

\bibitem{Liu19}
Z. X. Liu, H. Xiong, and Y. Wu,
Magnon blockade in a hybrid ferromagnet-superconductor quantum system,
Phys. Rev. B \textbf{100}, 134421 (2019).

\bibitem{Xie20}
J. K. Xie, S. L. Ma, and F. L. Li,
Quantum-interference-enhanced magnon blockade in an yttrium-iron-garnet sphere coupled to superconducting circuits,
Phys. Rev. A \textbf{101}, 042331 (2020).

\bibitem{Wang22-Xiong}
Y. Wang, W. Xiong, Z. Xu, G. Q. Zhang, and J. Q. You,
Dissipation-induced nonreciprocal magnon blockade in a magnon-based hybrid system,
Sci. China-Phys. Mech. Astron. \textbf{65}, 260314 (2022).

\bibitem{Harder17}
M. Harder, L. Bai, P. Hyde, and C. M. Hu,
Topological properties of a coupled spin-photon system induced by damping,
Phys. Rev. B \textbf{95}, 214411 (2017).

\bibitem{Cao19}
Y. Cao and P. Yan,
Exceptional magnetic sensitivity of $\mathcal{PT}$-symmetric cavity magnon polaritons,
Phys. Rev. B \textbf{99}, 214415 (2019).

\bibitem{Zhao20}
J. Zhao, Y. Liu, L. Wu, C. K. Duan, Y. Liu, and J. Du,
Observation of Anti-$\mathcal{PT}$-Symmetry Phase Transition in the Magnon-Cavity-Magnon Coupled System,
Phys. Rev. Appl. \textbf{13}, 014053 (2020).

\bibitem{Yao17}
B. Yao, Y. S. Gui, J. W. Rao, S. Kaur, X. S. Chen, W. Lu, Y. Xiao, H. Guo, K. P. Marzlin, and C. M. Hu,
Cooperative polariton dynamics in feedback-coupled cavities,
Nat. Commun. \textbf{8}, 1437 (2017).

\bibitem{Hei21}
X. L. Hei, X. L. Dong, J. Q. Chen, C. P. Shen, Y. F. Qiao, and P. B. Li,
Enhancing spin-photon coupling with a micromagnet,
Phys. Rev. A \textbf{103}, 043706 (2021).

\bibitem{Yuan20}
H. Y. Yuan, P. Yan, S. Zheng, Q. Y. He, K. Xia, and M.-H. Yung,
Steady Bell State Generation via Magnon-Photon Coupling,
Phys. Rev. Lett. \textbf{124}, 053602 (2020).

\bibitem{Sun21}
F. X. Sun, S. S. Zheng, Y. Xiao, Q. Gong, Q. He, and K. Xia,
Remote generation of magnon Schr\"{o}dinger cat state via magnon-photon entanglement,
Phys. Rev. Lett. \textbf{127}, 087203 (2021).

\bibitem{Zhang22}
G. Q. Zhang, W. Feng, W. Xiong, Q. P. Su, and C. P. Yang,
Generation of long-lived $W$ states via reservoir engineering in dissipatively coupled systems,
Phys. Rev. A \textbf{107}, 012410 (2023).

\bibitem{Qi22}
S. F. Qi and J. Jing,
Generation of Bell and Greenberger-Horne-Zeilinger states from a hybrid qubit-photon-magnon system,
Phys. Rev. A \textbf{105}, 022624 (2022).

\bibitem{Hisatomi16}
R. Hisatomi, A. Osada, Y. Tabuchi, T. Ishikawa, A. Noguchi, R. Yamazaki, K. Usami, and Y. Nakamura,
Bidirectional conversion between microwave and light via ferromagnetic magnons,
Phys. Rev. B \textbf{93}, 174427 (2016).

\bibitem{Zhu20}
N. Zhu, X. Zhang, X. Han, C. L. Zou, C. Zhong, C. H. Wang, L. Jiang, and H. X. Tang,
Waveguide cavity optomagnonics for broadband multimode microwave-to-optics conversion,
Optica \textbf{7}, 1291 (2020).

\bibitem{Grigoryan18}
V. L. Grigoryan, K. Shen, and K. Xia,
Synchronized spin-photon coupling in a microwave cavity,
Phys. Rev. B. \textbf{98}, 024406 (2018).

\bibitem{Harder18}
M. Harder, Y. Yang, B. M. Yao, C. H. Yu, J. W. Rao, Y. S. Gui, R. L. Stamps, and C. M. Hu,
Level Attraction Due to Dissipative Magnon-Photon Coupling,
Phys. Rev. Lett. \textbf{121}, 137203 (2018).

\bibitem{Zhang17}
D. Zhang, X. Q. Luo, Y. P. Wang, T. F. Li, and J. Q. You,
Observation of the exceptional point in cavity magnon-polaritons,
Nat. Commun. \textbf{8}, 1368 (2017).

\bibitem{You19}
G. Q. Zhang and J. Q. You,
Higher-order exceptional point in a cavity magnonics system,
Phys. Rev. B \textbf{99}, 054404 (2019).

\bibitem{Chong10}
Y. D. Chong, L. Ge, H. Cao, and A. D. Stone,
Coherent Perfect Absorbers: Time-Reversed Lasers,
Phys. Rev. Lett. \textbf{105}, 053901 (2010).

\bibitem{Wan11}
W. Wan, Y. Chong, L. Ge, H. Noh, A. D. Stone, and H. Cao,
Time-reversed lasing and interferometric control of absorption,
Science \textbf{331}, 889 (2011).

\bibitem{Sun14}
Y. Sun, W. Tan, H. Q. Li, J. Li, and H. Chen,
Experimental Demonstration of a Coherent Perfect Absorber with PT Phase Transition,
Phys. Rev. Lett. \textbf{112}, 143903 (2014).

\bibitem{Wang21}
C. Wang, W. R. Sweeney, A. D. Stone, and L. Yang,
Coherent perfect absorption at an exceptional point,
Science \textbf{373}, 1261 (2021).

\bibitem{Wong16}
Z. J. Wong, Y. L. Xu, J. Kim, K. O'Brien, Y. Wang, L. Feng, and X. Zhang,
Lasing and anti-lasing in a single cavity,
Nat. Photonics \textbf{10}, 796 (2016).

\bibitem{Pichler19}
K. Pichler, M. K\"{u}hmayer, J. B\"{o}hm, A. Brandst\"{o}tter, P. Ambichl, U. Kuhl, and S. Rotter,
Random anti-lasing through coherent perfect absorption in a disordered medium,
Nature (London) \textbf{567}, 351 (2019).

\bibitem{Fang14}
X. Fang, M. L. Tseng, J. Y. Ou, K. F. MacDonald, D. P. Tsai, and N. I. Zheludev,
Ultrafast all-optical switching via coherent modulation of metamaterial absorption,
Appl. Phys. Lett. \textbf{104}, 141102 (2014).

\bibitem{Xiong20-Chen}
W. Xiong, J. Chen, B. Fang, C. H. Lam, and J. Q. You,
Coherent perfect absorption in a weakly coupled atom-cavity system,
Phys. Rev. A \textbf{101}, 063822 (2020).

\bibitem{Kang15}
M. Kang and Y. D. Chong,
Coherent optical control of polarization with a critical metasurface,
Phys. Rev. A \textbf{92}, 043826 (2015).

\bibitem{Ye16}
Y. Ye, D. Hay, and Z. Shi,
Coherent perfect absorption in chiral metamaterials,
Opt. Lett. \textbf{41}, 3359 (2016).

\bibitem{Heiss12}
W. D. Heiss,
The physics of exceptional points,
J. Phys. A Math. Theor. \textbf{45}, 444016 (2012).

\bibitem{Gao15}
T. Gao, E. Estrecho, K. Y. Bliokh, T. C. H. Liew, M. D. Fraser, S. Brodbeck, M. Kamp, C. Schneider, S. H\"{o}fling, Y. Yamamoto, F. Nori, Y. S. Kivshar, A. G. Truscott, R. G. Dall, and E. A. Ostrovskaya,
Observation of non-Hermitian degeneracies in a chaotic exciton-polariton billiard,
Nature (London) \textbf{526}, 554 (2015).

\bibitem{Lv15}
X. Y. L\"{u}, H. Jing, J. Y. Ma, and Y. Wu,
$\mathcal{PT}$-Symmetry-Breaking Chaos in Optomechanics,
Phys. Rev. Lett. \textbf{114}, 253601 (2015).

\bibitem{Zhang21-Chen}
G. Q. Zhang, Z. Chen, D. Xu, N. Shammah, M. Liao, T. F. Li, L. Tong, S. Y. Zhu, F. Nori, and J. Q. You,
Exceptional point and cross-relaxation effect in a hybrid quantum system,
PRX Quantum \textbf{2}, 020307 (2021).

\bibitem{Jing17}
H. Jing, \c{S}. K. \"{O}zdemir, H. L\"{u}, and F. Nori,
High-order exceptional points in optomechanics,
Sci. Rep. \textbf{7}, 3386 (2017).

\bibitem{Lu21}
T. X. Lu, H. Zhang, Q. Zhang, and H. Jing,
Exceptional-point-engineered cavity magnomechanics,
Phys. Rev. A \textbf{103}, 063708 (2021).

\bibitem{Xiong22-Li}
W. Xiong, Z. Li, G. Q. Zhang, M. Wang, H. C. Li, X. Q. Luo, and J. Chen,
Higher-order exceptional point in a blue-detuned non-Hermitian cavity optomechanical system,
Phys. Rev. A \textbf{106}, 033518 (2022).

\bibitem{Doppler16}
J. Doppler, A. A. Mailybaev, J. B\"{o}hm, U. Kuhl, A. Girschik, F. Libisch, T. J. Milburn, P. Rabl, N. Moiseyev, and S. Rotter,
Dynamically encircling an exceptional point for asymmetric mode switching,
Nature (London) \textbf{537}, 76 (2016).

\bibitem{Zhang22-Liu}
J. Q. Zhang, J. X. Liu, H. L. Zhang, Z. R. Gong, S. Zhang, L. L. Yan, S. L. Su, H. Jing, and M. Feng,
Topological optomechanical amplifier in synthetic $\mathcal{PT}$-symmetry,
Nanophotonics \textbf{11}, 1149 (2022).

\bibitem{Zhiyenbayev19}
Y. Zhiyenbayev, Y. Kominis, C. Valagiannopoulos, V. Kovanis, and A. Bountis,
Enhanced stability, bistability, and exceptional points in saturable active photonic couplers,
Phys. Rev. A \textbf{100}, 043834 (2019).

\bibitem{Wiersig14}
J. Wiersig,
Enhancing the Sensitivity of Frequency and Energy Splitting Detection by Using Exceptional Points: Application to Microcavity Sensors for Single-Particle Detection,
Phys. Rev. Lett. \textbf{112}, 203901 (2014).

\bibitem{Chen17}
W. Chen, \c{S}. K. \"{O}zdemir, G. Zhao, J. Wiersig, and L. Yang,
Exceptional points enhance sensing in an optical microcavity,
Nature (London) \textbf{548}, 192 (2017).

\bibitem{Liu16}
Z. P. Liu, J. Zhang, \c{S}. K. \"{O}zdemir, B. Peng, H. Jing, X. Y. L\"{u}, C. W. Li, L. Yang, F. Nori, and Y. X. Liu,
Metrology with $\mathcal{PT}$-Symmetric Cavities: Enhanced Sensitivity near the $\mathcal{PT}$-Phase Transition,
Phys. Rev. Lett. \textbf{117}, 110802 (2016).

\bibitem{Zhang19-Wang-You}
G. Q. Zhang, Y. P. Wang, and J. Q. You,
Dispersive readout of a weakly coupled qubit via the parity-time-symmetric phase transition,
Phys. Rev. A \textbf{99}, 052341 (2019).

\bibitem{Wang21-Guo}
X. G. Wang, G. H. Guo, and J. Berakdar,
Enhanced Sensitivity at Magnetic High-Order Exceptional Points and Topological Energy Transfer in Magnonic Planar Waveguides,
Phys. Rev. Appl. \textbf{15}, 034050 (2021).

\bibitem{Wang16}
Y. P. Wang, G. Q. Zhang, D. Zhang, X. Q. Luo, W. Xiong, S. P. Wang, T. F. Li, C. M. Hu, and J. Q. You,
Magnon Kerr effect in a strongly coupled cavity-magnon system,
Phys. Rev. B \textbf{94}, 224410 (2016).

\bibitem{Zhang-China-19}
G. Q. Zhang, Y. P. Wang, and J. Q. You,
Theory of the magnon Kerr effect in cavity magnonics,
Sci. China-Phys. Mech. Astron. \textbf{62}, 987511 (2019).

\bibitem{Wang18}
Y. P. Wang, G. Q. Zhang, D. Zhang, T. F. Li, C. M. Hu, and J. Q. You,
Bistability of Cavity Magnon-Polaritons,
Phys. Rev. Lett. \textbf{120}, 057202 (2018).

\bibitem{Shen22}
R. C. Shen, J. Li, Z. Y. Fan, Y. P. Wang, and J. Q. You,
Mechanical Bistability in Kerr-modified Cavity Magnomechanics,
Phys. Rev. Lett. \textbf{129}, 123601 (2022).

\bibitem{Nair21}
J. M. P. Nair, D. Mukhopadhyay, and G. S. Agarwal,
Ultralow threshold bistability and generation of long-lived mode in a dissipatively coupled nonlinear system: Application to magnonics,
Phys. Rev. B \textbf{103}, 224401 (2021).

\bibitem{Shen21}
R. C. Shen, Y. P. Wang, J. Li, S. Y. Zhu, G. S. Agarwal, and J. Q. You,
Long-Time Memory and Ternary Logic Gate Using a Multistable Cavity Magnonic System,
Phys. Rev. Lett. \textbf{127}, 183202 (2021).

\bibitem{Nair20}
J. M. P. Nair, Z. Zhang, M. O. Scully, and G. S. Agarwal,
Nonlinear spin currents,
Phys. Rev. B \textbf{102}, 104415 (2020).

\bibitem{Bi21}
M. X. Bi, X. H. Yan, Y. Zhang, and Y. Xiao,
Tristability of cavity magnon polaritons,
Phys. Rev. B \textbf{103}, 104411 (2021).

\bibitem{Kong19}
C. Kong, H. Xiong, and Y. Wu,
Magnon-Induced Nonreciprocity Based on the Magnon Kerr Effect,
Phys. Rev. Appl. \textbf{12}, 034001 (2019).

\bibitem{Xiong22}
W. Xiong, M. Tian, G. Q. Zhang, and J. Q. You,
Strong long-range spin-spin coupling via a Kerr magnon interface,
Phys. Rev. B \textbf{105}, 245310 (2022).

\bibitem{Scully19}
Z. Zhang, M. O. Scully, and G. S. Agarwal,
Quantum entanglement between two magnon modes via Kerr nonlinearity driven far from equilibrium,
Phys. Rev. Research \textbf{1}, 023021 (2019).

\bibitem{Yang21}
Z. B. Yang, H. Jin, J. W. Jin, J. Y. Liu, H. Y. Liu, and R. C. Yang,
Bistability of squeezing and entanglement in cavity magnonics,
Phys. Rev. Research \textbf{3}, 023126 (2021).

\bibitem{Zhang21}
G. Q. Zhang, Z. Chen, W. Xiong, C. H. Lam, and J. Q. You,
Parity-symmetry-breaking quantum phase transition via parametric drive in a cavity magnonic system,
Phys. Rev. B \textbf{104}, 064423 (2021).

\bibitem{Qin22}
Y. Qin, S. C. Li, K. Li, and J. J. Song,
Controllable quantum phase transition in a double-cavity magnonic system,
Phys. Rev. B \textbf{106}, 054419 (2022).

\bibitem{Haigh15}
J. A. Haigh, N. J. Lambert, A. C. Doherty, and A. J. Ferguson,
Dispersive readout of ferromagnetic resonance for strongly coupled magnons and microwave photons,
Phys. Rev. B \textbf{91}, 104410 (2015).

\bibitem{Nair21-Mukhopadhyay}
J. M. P. Nair, D. Mukhopadhyay, and G. S. Agarwal,
Enhanced Sensing of Weak Anharmonicities through Coherences in Dissipatively Coupled Anti-PT Symmetric Systems,
Phys. Rev. Lett. \textbf{126}, 180401 (2021).

\bibitem{Gurevich96}
A. G. Gurevich and G. A. Melkov,
\emph{Magnetization Oscillations and Waves} (CRC Press, Boca Raton, 1996).

\bibitem{Stancil09}
D. D. Stancil and A. Prabhakar,
\emph{Spin Waves} (Springer, Berlin, 2009).

\bibitem{Walls94}
D. F. Walls and G. J. Milburn,
\emph{Quantum Optics} (Springer, Berlin, 1994).

\bibitem{Mostafazadeh021}
A. Mostafazadeh,
Pseudo-Hermiticity versus $\mathcal{PT}$-symmetry: The necessary condition for the reality of the spectrum of a non-Hermitian Hamiltonian,
J. Math. Phys. \textbf{43}, 205 (2002).

\bibitem{Mostafazadeh022}
A. Mostafazadeh,
Pseudo-Hermiticity versus $\mathcal{PT}$-symmetry II: A complete characterization of non-Hermitian Hamiltonians with a real spectrum,
J. Math. Phys. \textbf{43}, 2814 (2002).

\bibitem{Mostafazadeh023}
A. Mostafazadeh,
Pseudo-Hermiticity versus $\mathcal{PT}$-symmetry III: Equivalence of pseudo-Hermiticity and the presence of antilinear symmetries,
J. Math. Phys. \textbf{43}, 3944 (2002).

\bibitem{Hodaei17}
H. Hodaei, A. U. Hassan, S. Wittek, H. Garcia-Gracia, R. El-Ganainy, D. N. Christodoulides, and M. Khajavikhan,
Enhanced sensitivity at higher-order exceptional points,
Nature (London) \textbf{548}, 187 (2017).

\bibitem{Xiong21}
W. Xiong, Z. Li, Y. Song, J. Chen, G. Q. Zhang, and M. Wang,
Higher-order exceptional point in a pseudo-Hermitian cavity optomechanical system,
Phys. Rev. A \textbf{104}, 063508 (2021).

\bibitem{Zeng21}
C. Zeng, K. Zhu, Y. Sun, G. Li, Z. Guo, J. Jiang, Y. Li, H. Jiang, Y. Yang, and H. Chen,
Ultra-sensitive passive wireless sensor exploiting high-order exceptional point for weakly coupling detection,
New J. Phys. \textbf{23}, 063008 (2021).

\bibitem{Wiersig20}
J. Wiersig,
Review of exceptional point-based sensors,
Photonics Res. \textbf{8}, 1457 (2020).

\bibitem{Wolff19}
C. Wolff, C. Tserkezis, and N. A. Mortensen,
On the time evolution at a fluctuating exceptional point,
Nanophotonics \textbf{8}, 1319 (2019).

\bibitem{Wiersig20-PRA}
J. Wiersig,
Robustness of exceptional-point-based sensors against parametric noise: the role of Hamiltonian and Liouvillian degeneracies,
Phys. Rev. A \textbf{101}, 053846 (2020).

\bibitem{Langbein18}
W. Langbein,
No exceptional precision of exceptional-point sensors,
Phys. Rev. A \textbf{98}, 023805 (2018).

\bibitem{Lau18}
H. K. Lau and A. A. Clerk,
Fundamental limits and non-reciprocal approaches in non-Hermitian quantum sensing,
Nat. Commun. \textbf{9}, 4320 (2018).

\bibitem{Chen-Jin19}
C. Chen, L. Jin, and R. B. Liu,
Sensitivity of parameter estimation near the exceptional point of a non-Hermitian system,
New J. Phys. \textbf{21}, 083002 (2019).

\bibitem{Zhang-Sweeney19}
M. Zhang, W. Sweeney, C. W. Hsu, L. Yang, A. D. Stone, and L. Jiang,
Quantum noise theory of exceptional point amplifying sensors,
Phys. Rev. Lett. \textbf{123}, 180501 (2019).

\bibitem{Kruse16}
I. Kruse, K. Lange, J. Peise, B. L\"{u}cke, L. Pezz\`{e}, J. Arlt, W. Ertmer, C. Lisdat, L. Santos, A. Smerzi, and C. Klempt,
Improvement of an Atomic Clock using Squeezed Vacuum,
Phys. Rev. Lett. \textbf{117}, 143004 (2016).

\bibitem{Malnou19}
M. Malnou, D. A. Palken, B. M. Brubaker, L. R. Vale, G. C. Hilton, and K. W. Lehnert,
Squeezed Vacuum Used to Accelerate the Search for a Weak Classical Signal,
Phys. Rev. X \textbf{9}, 021023 (2019).

\end{thebibliography}
\end{document}